\documentclass[prd,preprint,superscriptaddress,amsmath,amssymb,nofootinbib]{revtex4}
\usepackage{graphicx}
\usepackage{dcolumn}
\usepackage{bm}
\usepackage{amssymb}
\usepackage{amsmath}
\usepackage{epsfig}    
\usepackage{color}
\usepackage{slashed}
\usepackage{hhline}

\def\be{\begin{equation}}
\def\ee{\end{equation}}
\newcommand{\bea}{\begin{eqnarray}}
\newcommand{\eea}{\end{eqnarray}}

\begin{document}

 \begin{flushright} {CTP-SCU/2021020}, APCTP Pre2021-014 \end{flushright}

\title{
A Radiative Neutrino Mass Model \\
in Dark Non-Abelian Gauge Symmetry}
\author{Takaaki Nomura}
\email{nomura@scu.edu.cn}
\affiliation{College of Physics, Sichuan University, Chengdu 610065, China}

\author{Hiroshi Okada}
\email{hiroshi.okada@apctp.org}
\affiliation{Asia Pacific Center for Theoretical Physics, Pohang 37673, Republic of Korea}
\affiliation{Department of Physics, Pohang University of Science and Technology, Pohang 37673, Republic of Korea}

\date{\today}

\begin{abstract}
We discuss a model based on dark sector described by non-Abelian $SU(2)_D$ gauge symmetry 
where we introduce $SU(2)_L \times SU(2)_D$ bi-doublet vector-like leptons to generate active neutrino masses 
and kinetic mixing between $SU(2)_D$ and $U(1)_Y$ gauge fields at one-loop level.
After spontaneous symmetry breaking of $SU(2)_D$, we have remnant $Z_4$ symmetry guaranteeing stability of dark matter candidates.
We formulate neutrino mass matrix and related lepton flavor violating processes and discus dark matter physics estimating relic density.
It is found that our model realize multicomponent dark matter scenario due to the $Z_4$ symmetry 
and relic density can be explained by gauge interactions with kinetic mixing effect.

 \end{abstract}
\maketitle

\section{Introduction}

A mechanism of generating neutrino mass and the existence of dark matter (DM) are important hints to understand physics beyond the standard model(SM).
One of the attractive scenarios is that DM and neutrino mass generation are induced from dark sector described by dark gauge symmetry under which the SM fields are singlet.
Then we expect nature of neutrino mass generation mechanism and DM physics are understand by dark gauge symmetry.
For example, the stability of DM could be understood by remnant of dark gauge symmetry~\cite{Krauss:1988zc, Ko:2018qxz} and neutrino mass at tree level can be suppressed by such symmetry.

One interesting scenario is non-Abelian dark gauge symmetry such as $SU(2)$ which provides us rich structure of dark sector, giving 
possibility of vector DM from dark gauge sector and $Z'$ as mediator at the same time.
In fact, we can find various approaches applying a dark $SU(2)$ gauge symmetry in literatures, for examples, a remaining $Z_{2,3,4}$ symmetry with a quadruplet(quintet) in ref.~\cite{Chiang:2013kqa, Chen:2015nea,Chen:2015dea,Chen:2015cqa,Ko:2020qlt,Nomura:2020zlm,Chen:2017tva}, $Z_2 \times Z'_2$ symmetry~\cite{Gross:2015cwa}, a custodial symmetry in refs.~\cite{Boehm:2014bia, Hambye:2008bq,Baouche:2021wwa}, an unbroken $U(1)$ from $SU(2)$ in refs.~\cite{Baek:2013dwa,Khoze:2014woa,Daido:2019tbm}, a model adding hidden $U(1)_h$~\cite{Davoudiasl:2013jma}, other DM scenarios~\cite{Barman:2017yzr,Barman:2018esi,Barman:2019lvm,Barman:2020ifq}, a model with classical scale invariance~\cite{Karam:2015jta}, Baryogengesis~\cite{Hall:2019ank} and electroweak phase transition~\cite{Ghosh:2020ipy}.
Here one interesting question for non-Abelian dark gauge symmetric case is how we can induce interactions among dark gauge bosons and the SM particles,
since kinetic mixing is not allowed at renormalizable level in contrast to the Abelian gauge symmetric case.
In ref.~\cite{Nomura:2021tmi}, we showed one-loop generation of a term generating kinetic mixing between dark $SU(2)$ and the SM $U(1)_Y$ introducing a field which has both dark $SU(2)$ and $U(1)_Y$ charge.
Interestingly when we chose such a field as vector-like leptons, they can also play a role in generating active neutrino mass at loop level~\cite{Okada:2013iba, Okada:2014qsa,Okada:2015vwh} by adding relevant dark $SU(2)$ multiplet fields.

In this work, we discuss a model with non-Abelian $SU(2)_D$ gauge symmetry in which we introduce $SU(2)_L \times SU(2)_D$ bi-doublet vector like leptons.
This bi-doublet leptons can induce mixing among $SU(2)_D$ and $U(1)_Y$ gauge fields and play a role to generate active neutrino mass
when we introduce relevant scalar $SU(2)_D$ multiplets.
It is also found that there is remnant $Z_4$ symmetry after spontaneous symmetry breaking and stability of DM is guaranteed by this symmetry.
We then formulate active neutrino mass and branching rations (BRs) of lepton flavor violating (LFV) charged lepton decay $\ell_i \to \ell_j \gamma$ in our model.
In addition, relic density of our DM candidates is estimated where DM is more than one component in our scenario.

This paper is organized as follows.
In Sec.II, we introduce our model showing relevant Lagrangian and particle contents.
In Sec.III, we discuss phenomenology of the model such as neutrino mass generation, LFV and DM physics.
Summary and discussion are given in Sec.IV.

 \begin{widetext}
\begin{center} 
\begin{table}[t]
\begin{tabular}{|c||c|c||c|c|c|}\hline\hline  
Fields  & $L'$ & $N$& $\chi$ & $\varphi$ & $\Phi$
\\\hline 
 $SU(2)_D$ &  $\bf{2}$ & $\bf{2}$ &  $\bf{2}$ & $\bf{3}$ & $\bf{5}$    \\\hline 
 $SU(2)_L$  & $\bf{2}$  & $\bf{1}$ & $\bf{1}$  & $\bf{1}$ & $\bf{1}$   \\\hline
 $U(1)_Y$ & $-\frac12$ & $0$ & $0$ & $0$ & $0$   \\\hline
\end{tabular}
\caption{Charge assignment for the fields in $SU(2)_D$ dark sector where $\{\chi, \varphi, \Phi \}$ are scalars and $\{ L', N\}$ are Dirac fermions.}
\label{tab:1}
\end{table}
\end{center}
\end{widetext}

\section{A model}
We consider a model based on $G_{SM} \times SU(2)_D$ gauge symmetry where $G_{SM}$ is the SM gauge symmetry and $SU(2)_D$ is additional one in our dark sector.
For fermion sector, we introduce $SU(2)_L \times SU(2)_D$ bi-doublet lepton $L'$ with $U(1)_Y$ charge $-1/2$ and $SU(2)_D$ doublet $N$ which is singlet under $G_{SM}$.
Here three generations of these fermions are considered in our model.
For scalar sector, we introduce $SU(2)_D$ complex quintet $\Phi$,  real triplet $\varphi$ and complex doublet $\chi$; the SM Higgs doublet $H$ is also included. 
The new field contents are summarized in Table~\ref{tab:1} with their charge assignments.
We write $L'$ and $N$ by 
\begin{equation}
L' = \begin{pmatrix} n'_{1} & n'_{2} \\ e'_1 & e'_2 \end{pmatrix}, \quad N = \begin{pmatrix} n_1 \\ n_2 \end{pmatrix},
\end{equation}
where indices for generation are omitted.
The scalar multiplets are also written by 
\begin{align}
\label{eq:multiplet}
\chi = \begin{pmatrix} \chi_1 \\ \chi_2 \end{pmatrix}, \quad \varphi = \begin{pmatrix} \frac{\varphi_0}{\sqrt{2}} & \varphi_+ \\ \varphi_- & - \frac{\varphi_0}{\sqrt{2}} \end{pmatrix}, \quad 
\Phi = \begin{pmatrix} \Phi_{++} & \Phi_{+} & \Phi_0 & \tilde \Phi_{-} & \tilde \Phi_{--} \end{pmatrix}^T,
\end{align}
where $\varphi_+ = (\varphi_-)^*$. 
The triplet $\varphi$ can be written by $\varphi^\alpha \sigma_\alpha/\sqrt{2} \ (\alpha=1,2,3)$ with $\sigma^\alpha$ being the Pauli matrix acting on $SU(2)_D$ representation space;
thus we define $\varphi_0 = \varphi^3$ and $\varphi_\pm = (\varphi^1 \mp i \varphi^2)/\sqrt{2}$.
The SM Higgs field is written by
\begin{equation}
H = \begin{pmatrix} G^+ \\ \frac{1}{\sqrt{2}} (v + h  + i G^0) \end{pmatrix},
\end{equation}
where $v \simeq 246$ GeV {is vacuum expectation value (VEV)} and $G^{+(0)}$ is Nambu-Goldstone(NG) boson absorbed by $W^+(Z)$ boson. 

The Lagrangian of our model is written by
\begin{align}
\mathcal{L} = \mathcal{L}_{\rm SM} + \mathcal{L}_{\rm New} + V,
\end{align}
where $\mathcal{L}_{SM}$ is the SM Lagrangian without Higgs potential, $\mathcal{L}_{\rm New}$ includes new terms in our model and $V$ is the scalar potential.
The new terms and the potential are given such that
\begin{align}
\label{eq:Lnew}
 \mathcal{L}_{\rm New} = & -\frac14 \tilde X^{\alpha \mu \nu} \tilde X^\alpha_{\mu \nu} + {\rm Tr}[\bar L' (D_\mu \gamma^\mu - M_{L'}) L'] + \bar N (\partial^\mu \gamma_\mu - M_N) N \nonumber \\
& +  (D^\mu \chi)^\dagger (D_\mu \chi) + \frac12 {\rm Tr}[(D^\mu \varphi)^\dagger (D_\mu \varphi)] + \frac12 (D^\mu \Phi)^\dagger (D_\mu \Phi) \nonumber \\
 & +f_{ia} \bar L^i_L L'^a_R (i \sigma_2) \chi + f'_{ia} \bar L^i_L L'^a_R  \chi^* + g_R^{ab} \bar L'^a_R N^b_L \tilde H + g_L^{ab} \bar L'^a_L N^b_R \tilde H \nonumber \\
& + y_{N_L}^{ab}  \overline{N_L^{a c}} (i \sigma_2) \varphi N^b_L + y_{N_R}^{ab}  \overline{N_R^{a c}} (i \sigma_2) \varphi N^b_L + y_D^{ab} \bar N^a_L \varphi N^b_R
+ y_{ab} \bar L'^a \varphi L'^b  \\
\label{eq:potential1}
 V = & \ - M_H^2 H^\dagger H  + M_\chi^2 \chi^\dagger \chi  + \frac12 M_\varphi^2 {\rm Tr}[\varphi \varphi] - M_\Phi^2 \Phi^\dagger \Phi + \lambda_\chi (\chi^\dagger \chi)^2 + \lambda_\varphi {\rm Tr}[\varphi \varphi]^2 + \lambda_\Phi (\Phi^\dagger \Phi)^2 \nonumber  \\
 & + \lambda_H (H^\dagger H)^2+ \mu_1 (\Phi^\dagger \hat \varphi \Phi) + \mu_2 ( \chi (i \sigma_2) \varphi \chi +h.c.) + \lambda_{\varphi \Phi} {\rm Tr}[ \varphi \varphi] (\Phi^\dagger \Phi)  \nonumber \\
 &  + \lambda_{\varphi H}  {\rm Tr}[ \varphi \varphi] (H^\dagger H)+ \lambda_{\Phi H} (\Phi^\dagger \Phi) (H^\dagger H)   + \lambda_{\chi H} (\chi^\dagger \chi)( H^\dagger H) + \lambda_{\chi \varphi} (\chi^\dagger \chi) {\rm Tr}[ \varphi \varphi] 
 \nonumber \\
 & + \lambda_{\chi \Phi} (\chi^\dagger \chi) (\Phi^\dagger \Phi)+ \tilde \lambda_{\varphi \Phi} \Phi^\dagger \hat \varphi \hat \varphi \Phi,
\end{align}
where $\sigma_2$ is the second Pauli matrix acting on $SU(2)_D$ representation space, $\tilde X^{\alpha \mu \nu}$ is the gauge field strength for $SU(2)_D$ with $\alpha = 1,2,3$ being index of $SU(2)_D$ adjoint representation,
and $\hat \varphi \equiv \varphi^\alpha \mathcal{T}^{(5)}_\alpha$ is $5 \times 5$ notation of scalar triplet ($\mathcal{T}^{(5)}_\alpha$ is $5\times 5$ notation of $SU(2)_D$ generation given in the Appendix).
We assume Lagrangian is invariant under $\Phi \to - \Phi$ to simplify scalar potential forbidding non-trivial cubic terms such as $\Phi^3$ and $\Phi \varphi \varphi$.

\subsection{Scalar sector and symmetry breaking}

Firstly we consider gauge invariant operators in scalar potential in terms of the components of the scalar multiplets. 
Quadratic terms are given by
\begin{align}
& \chi^\dagger \chi = \chi_1^* \chi_1 + \chi_2^* \chi_2, \\
& \frac12 {\rm Tr}[\varphi \varphi] = \frac12 \varphi_0^2 + \varphi_+ \varphi_-, \\
&  \Phi^\dagger \Phi =  \Phi_0^2 + \Phi_{++} \Phi_{--} + \Phi_{+} \Phi_{-} + \tilde \Phi_{++} \tilde \Phi_{--} + \tilde \Phi_{+} \tilde \Phi_{-}.
\end{align}
Non-trivial terms in the potential are written by
\begin{align}
& \Phi^\dagger \hat \varphi \Phi =  \sqrt{3} \Phi_0 \tilde \Phi_+ \varphi_- + \sqrt{3} \Phi_0  \Phi_- \varphi_+ + \sqrt{3} \Phi_0^* \Phi_+ \varphi_- + \sqrt{3} \Phi_0^* \tilde \Phi_- \varphi_+  \nonumber \\
& \qquad \qquad + (2 \Phi_{++} \Phi_{--} + \Phi_{+} \Phi_{-} - \tilde \Phi_{+} \tilde \Phi_{-} - 2 \tilde \Phi_{++} \tilde \Phi_{--})  \varphi_0 \nonumber \\
& \qquad \qquad +\sqrt{2}  \Phi_{--} \Phi_{+} \varphi_+ + \sqrt{2} \tilde \Phi_{--} \tilde \Phi_+ \varphi_+ + \sqrt{2} \Phi_{++} \Phi_- \varphi_- + \sqrt{2} \tilde \Phi_{++} \tilde \Phi_- \varphi_-   \\
& \Phi^\dagger \hat \varphi \hat \varphi \Phi = \varphi_0^2 (4 \Phi_{++} \Phi_{--} + 4 \tilde \Phi_{++} \tilde \Phi_{--} + \Phi_{+} \Phi_{-} +  \tilde \Phi_{+} \tilde \Phi_{-}) \nonumber \\
& \qquad \qquad + \varphi_+ \varphi_- (2 \Phi_{++} \Phi_{--} + 2 \tilde \Phi_{++} \tilde \Phi_{--} + 5 \Phi_+ \Phi_- + 5 \tilde \Phi_+ \tilde \Phi_- ) \nonumber \\
& \qquad \qquad + \varphi_-^2 (3 \Phi_+ \tilde \Phi_+ + \sqrt{6} \Phi_0 \tilde \Phi_{++} + \sqrt{6} \Phi_0^* \Phi_{++}) 
+ \varphi_+^2 (3 \Phi_- \tilde \Phi_- + \sqrt{6} \Phi^*_0 \tilde \Phi_{--} + \sqrt{6} \Phi_0 \Phi_{--}) \nonumber \\
& \qquad \qquad + \sqrt{3} \varphi_0 \Phi_0 (\Phi_- \varphi_+ - \tilde \Phi_+ \varphi_-) + \sqrt{3} \varphi_0 \Phi^*_0 (\Phi_+ \varphi_- - \tilde \Phi_- \varphi_+) \nonumber \\
& \qquad \qquad + 3 \sqrt{2} \varphi_0 (\Phi_{++} \Phi_- \varphi_- + \Phi_{--} \Phi_+ \varphi_+ - \tilde \Phi_{++} \tilde \Phi_- \varphi_- - \tilde \Phi_{--} \tilde \Phi_+ \varphi_+ ), \\
&  \chi (i \sigma_2) \varphi \chi +h.c. = - \sqrt{2} \varphi_0 \chi_1 \chi_2 + \varphi_- \chi_1 \chi_1 - \varphi_+ \chi_2 \chi_2 + h.c. \, .
\end{align}
Note that the other quartet terms are trivially given by applying quadratic terms and we do not write them explicitly.
We then consider VEVs of the scalar fields by the conditions $\partial V/\partial \phi =0$ where $\phi$ represents any scalar field in the model.
It is found that we can take VEVs of $\tilde \Phi_{\pm \pm}$ and $\varphi_0$ to be non-zero and we write them by $\langle \tilde \Phi_{\pm \pm} \rangle \equiv v_\Phi/\sqrt{2}$ and $\langle \varphi_{0} \rangle \equiv v_\varphi/\sqrt{2}$.
These VEVs are derived from the following conditions
\begin{align}
\label{eq:VEV1}
&  v_\Phi \left( -M_\Phi^2 -  \sqrt{2} \mu_1 v_\varphi +2 \tilde \lambda_{\varphi \Phi} v_\varphi^2 + \frac12 \lambda_{\Phi H} v^2 + \frac12 \lambda_{\varphi \Phi} v_\varphi^2  + \lambda_\Phi v_\Phi^2 \right)  = 0 \\
\label{eq:VEV2}
& - \frac{\mu_1}{\sqrt{2}} v_\Phi^2 + 2 \tilde \lambda_{\varphi \Phi} v_\varphi v_\Phi^2 +  M_\varphi^2 v_\varphi  + \frac12 \lambda_{\varphi H} v_\varphi v^2+ \frac12 \lambda_{\varphi \Phi} v_\varphi v_\Phi^2 + \lambda_\varphi v_\varphi^3 = 0 \\
\label{eq:VEV3}
& v \left(- M_H^2 + \frac12 \lambda_{\varphi H} v_\varphi^2 + \frac12 \lambda_{\Phi H} v_\Phi^2 + \lambda_H v^2 \right) =0,
\end{align}
where the first, second and third equations are obtained from $\partial V/\partial v_\Phi =0$, $\partial V/\partial v_\varphi =0$ and $\partial V/\partial v =0$ respectively.
In our analysis we consider mixing among the SM Higgs and other scalars are suppressed by assuming tiny values for $\lambda_{\varphi H}$ and $\lambda_{\Phi H}$,
and the SM Higgs VEV is approximately given by $v \simeq \sqrt{M_H^2/\lambda_H}$ as in the SM; thus $h$ is the SM-like Higgs boson.
On the other hand VEVs $v_\varphi$ and $v_\Phi$ are determined by Eqs.~\eqref{eq:VEV1} and \eqref{eq:VEV2}.
After spontaneous symmetry breaking, three degrees of freedom from $\Phi$ and $\varphi$ are absorbed by $SU(2)_D$ gauge bosons $\tilde X^{1,2,3}_\mu$ and 
the remaining scalar degrees of freedom from $\varphi$ and $\Phi$ become massive physical scalar bosons. 
Here we do not discuss much details of these physical scalar bosons since they are irrelevant in our phenomenological analysis below.
Note that there is remaining $Z_4$ symmetry in our scenario where each $SU(2)_D$ multiplet $\xi$ transform as $\xi \to e^{i T_3 \pi} \xi$ with $T_3$ 
being diagonal $SU(2)_D$ generator. 
Thus components in $SU(2)_D$ doublet have $Z_4$ charge, $\pm i$ while components of triplet and quintet have charge $\pm1$.
As a result $Z_4$ charged particles can be stable and become our DM candidates.

The scalar bosons from $\chi$ will be one of our DM candidate since they transform as $\chi_1 (\chi_2) \to i \chi_1(-i \chi_2)$ under remnant $Z_4$ symmetry.
The mass matrix for $\chi$ is obtained as 
\begin{equation}
\begin{pmatrix} \chi_1^* \\ \tilde{\chi}_2^* \end{pmatrix}^T
\begin{pmatrix} \tilde M^2_\chi & - \sqrt{2} \mu_2 v_\varphi \\ - \sqrt{2} \mu_2 v_\varphi & \tilde M^2_\chi \end{pmatrix}
\begin{pmatrix} \chi_1 \\ \tilde{\chi}_2 \end{pmatrix},
\end{equation}
where $\tilde \chi_2 \equiv \chi_2^*$ and $\tilde M^2_\chi = M^2_\chi + \frac12 \lambda_{\chi H} v^2 + \frac12 \lambda_{\chi \varphi} v_\varphi^2 +  \frac12 \lambda_{\chi \Phi} v_\Phi^2$.
We then obtain mass eigenstates and eigenvalues such that
\begin{align}
& \begin{pmatrix} \chi_1 \\ \tilde{\chi}_2 \end{pmatrix}^T = 
\begin{pmatrix} \frac{1}{\sqrt{2}} & -\frac{1}{\sqrt{2}} \\ \frac{1}{\sqrt{2}} & \frac{1}{\sqrt{2}} \end{pmatrix}
\begin{pmatrix} \rho_1 \\ \rho_2 \end{pmatrix}, \\
& m^2_{1,2} = \tilde M^2_\chi \pm \sqrt{2} \mu_2 v_\varphi
\end{align}
where we denote $m_{1,2}$ as masses of $\rho_{1,2}$ choosing $m_{1} < m_{2}$ and assume $\tilde M^2_\chi > \sqrt{2} \mu_2 v_\varphi$ to make eigenvalues positive.

\subsection{Gauge sector}
Here we focus on gauge sector of $SU(2)_L \times SU(2)_D \times U(1)_Y$ where the Lagrangian is
\begin{equation}
\mathcal{L}_G = - \frac14 W^{a \mu \nu} W^a_{\mu \nu} -\frac14 \tilde X^{\alpha \mu \nu} \tilde X^\alpha_{\mu \nu} - \frac14 \tilde B_{\mu \nu} \tilde B^{\mu \nu}, 
\label{eq:gauge1}
\end{equation}
and $W^{a \mu \nu}$ and $\tilde B^{\mu \nu}$ are gauge field strength of $SU(2)_L$ and $U(1)_Y$, respectively.
In addition to these terms in Eq.~\eqref{eq:gauge1}, a term connecting $SU(2)_D$ and $U(1)_B$ is generated by a one-loop diagrams 
in which $L'$ propagates~\cite{Nomura:2021tmi}.
We obtain such a term as 
\begin{equation}
\mathcal{L}_{XB} =  \sum_a \frac{g_X g_B y_{aa}}{12 \pi^2 M_{L'} } \tilde B_{\mu \nu} \tilde X^{\alpha \mu \nu} \varphi^\alpha.
\end{equation}
Then after $\varphi$ developing its VEV, we obtain kinetic mixing term 
\begin{equation}
\mathcal{L}_{KM} =  \sum_a \frac{g_X g_B y_{aa} v_\varphi}{12 \sqrt{2} \pi^2 M_{L'} } \tilde B_{\mu \nu} \tilde X^{3\mu\nu} \equiv -\frac12 \sin \chi \tilde B_{\mu \nu} \tilde X^{3 \mu \nu},
\end{equation}
where $\tilde X^{3}_{\mu \nu} \equiv \partial_\mu \tilde X^3_\nu - \partial_\nu \tilde X^3_\mu$.
We thus find magnitude of kinetic mixing parameter $\sin \chi$ as
\begin{equation}
\sin \chi \simeq 1.8 \times 10^{-3}  \frac{g_X}{0.5} \frac{v_\varphi}{M_{L'}} \sum_a y_{aa}.
\end{equation}
In our analysis we consider $\sin \chi < 10^{-3}$.
We can diagonalize the kinetic terms for $\tilde X^{3}_\mu$ and $\tilde{B}_\mu$ by the following transformations: 
\begin{align}
& \tilde{B}_\mu = B_\mu - \tan { \chi} X^3_\mu , \\
& \tilde{X}^3_\mu = \frac{1}{\cos { \chi} } X^3_\mu.
\end{align}
Since the kinetic mixing term is generated at loop level, $\chi$ is typically very small, and we take a limit of ${\chi} \ll 1$ writing gauge field approximately by
\begin{equation}
\tilde{B}_\mu \simeq B_\mu - { \chi} X^3_\mu ,  \quad \tilde{X}^3_\mu \simeq X^3_\mu.
\end{equation}
After quintet and triplet scalar fields develop nonzero VEVs, mass terms for $SU(2)_D$ gauge fields and SM Z boson field are given such that  
\begin{align}
& L_{M} = \frac{1}{2} m_{Z_{SM}}^2 \tilde{Z}_\mu \tilde{Z} + m_{Z_{SM}}^2  \chi \sin \theta_W \tilde Z_{\mu} X^{3 \mu} + \frac{1}{2} m_{X^3}^2 X^{3}_\mu X^{3 \mu} + m_{X^\pm}^2 X^+_\mu X^{- \mu}, \\
& m_{Z_{SM}}^2 = \frac{v^2}{4} (g^2+g_B^2), \quad m_{X^3}^2 = 4 g_X^2 v_\Phi^2, \quad m_{X^\pm}^2 = g_X^2 v_\Phi^2 \left(1+\frac{ v_\varphi^2}{v_\Phi^2}\right),
\end{align}
where $g$ and $g_B$ are gauge couplings of $SU(2)$ and $U(1)_{Y}$, $\tilde Z$ is $Z$ boson field in the SM, and $X_\mu^\pm = (\tilde X_\mu^1 \mp i X_\mu^2)/\sqrt{2}$.
Diagonalizing $\tilde Z$ and $X^3$ mass terms, we obtain mass eigenstate and mixing angles as
\begin{align}
& m_{Z, Z'}^2 = \frac{1}{2} (m_{X^3}^2 + m_{Z_{SM}}^2 ) \mp  \frac{1}{2} \sqrt{(m_{X^3}^2- m_{Z_{SM}}^2)^2 + 4 \chi^2 \sin^2 \theta_W m_{Z_{SM}}^4}, \\
& \tan 2 \theta_{ZZ'} =  \frac{2 \sin \theta_W \chi m_{Z_{SM}}^2}{m_{Z_{SM}}^2 - m_{X^3}^2},
\end{align}
where we approximate $m_Z \simeq m_{Z_{SM}}$ and $m_{Z'} \simeq m_{X^3}$ for tiny $\chi$, and we take $0 \leq \theta_{ZZ'} \leq \frac{\pi}{2}$ as our convention.
Here we emphasize that mass relation $m_{Z'} \sim 2m_{X^\pm}$ is obtained when dark gauge boson masses are dominantly induced by quintet VEV,
and $X^\pm$ annihilation cross section via $Z'$ is enhanced by resonant effect.

Thus the mass eigenstates $Z$ and $Z'$ are written by 
\begin{equation}
\begin{pmatrix} Z \\ Z' \end{pmatrix} = \begin{pmatrix} \cos \theta_{ZZ'} & - \sin \theta_{ZZ'} \\  \sin \theta_{ZZ'} & \cos \theta_{ZZ'} \end{pmatrix} \begin{pmatrix} \tilde Z \\ X^3 \end{pmatrix}.
\end{equation}
We find that $\sin \theta_{ZZ'} \lesssim 10^{-4}$ for $\chi \leq 10^{-3}$ and $m_{X^\pm} \geq 100$ GeV, and thus we can easily avoid current constraints~\cite{Langacker:2008yv}.

\subsection{Mass terms of hidden fermions}

In this subsection we discuss mass terms for hidden fermions.
{\it Neutral fermion masses:}
After spontaneous symmetry breaking we obtain mass terms of dark neutral fermions such that
\begin{align}
L_{M_N} = & M_N^{ab} (\overline{ n_1^a} n_1^b + \overline{ n_2^a} n_2^b) + M_{nn}^{ab} (\overline{n^{c a}_{2L}} n^b_{1L}+ \overline{n^{c a}_{1L}} n^b_{2L}) + \tilde{M}_{nn}^{ab} (\overline{n^{c a}_{2R}} n^b_{1R}+ \overline{n^{c a}_{1R}} n^b_{2R}) \nonumber \\
& + M_{D_n}^{ab} (\overline{n^a_{1L}} n^b_{1R} - \overline{n^a_{2L}} n^b_{2R} ) + M_{nn'}^{ab} (\overline{n'^a_{1R}} n^b_{1L} + \overline{n'^a_{2R}} n^b_{2L}) + \tilde{M}_{nn'}^{ab} (\overline{n'^a_{1L}} n^b_{1R} + \overline{n'^a_{2L}} n^b_{2R}) \nonumber \\
& + M_{L'}^{ab} (\overline{ n'^a_1} n'^b_1 + \overline{ n'^a_2} n'^b_2),
\end{align}
where mass matrices appearing these terms are given by
\begin{align}
M_{nn'}^{ab} = \frac{g_R^{ab}}{\sqrt{2}} v_H, \quad \tilde{M}_{nn'}^{ab} = \frac{g_L^{ab}}{\sqrt{2}} v_H, \quad M_{nn}^{ab} = \frac{y_{N_L}^{ab}}{2} v_\varphi, \quad \tilde{M}_{nn}^{ab} = \frac{y_{N_R}^{ab}}{2} v_\varphi,
\quad M_{D_n} = \frac{y_D^{ab}}{2} v_\varphi.
\end{align}
We thus obtain Majorana mass matrix for dark neutral fermions under the basis of $\Psi_R = (n'_{1R}, n'_{2R}, n'^c_{1L}, n'^c_{2L},n_{1R}, n_{2R}, n^c_{1L}, n^c_{2L})^T$ such as
\begin{equation}
M_R = \begin{pmatrix} 
0 & 0 & M_{L'}^T & 0 & 0 & 0 & M^*_{nn'} & 0 \\
0 & 0 & 0 & M_{L'}^T & 0 & 0 & 0 & M^*_{nn'} \\
M_{L'} & 0 & 0 & 0 & \tilde{M}_{nn'} & 0 & 0 & 0 \\
0 & M_{L'} & 0 & 0 & 0 & \tilde{M}_{nn'} & 0 & 0 \\
0 & 0 & \tilde{M}^T_{nn'} & 0 & 0 & \tilde{M}_{nn}^\dagger  & \tilde{M}^T_{N}+M^T_{D_n} & 0 \\
0 & 0 & 0 & \tilde{M}^T_{nn'} & \tilde{M}_{nn}^\dagger & 0  & 0 & \tilde{M}^T_{N}-M^T_{D_n} \\
M^\dagger_{nn'} & 0 & 0 & 0 & \tilde{M}_{N}+M_{D_n} & 0  & 0 & M_{nn}^\dagger  \\
0 & M^\dagger_{nn'} & 0 & 0 & 0 & \tilde{M}_{N}-M_{D_n}  & M_{nn}^\dagger & 0  \\
\end{pmatrix} , 
\end{equation}
where generation indices are omitted and it is $24 \times 24$ matrix including generation.
This mass matrix can be diagonalized by $24 \times 24$ orthogonal matrix $V_N$ assuming all matrix elements are real, and mass eigenstates are given by
\begin{equation}
\psi_R = V_N^T \Psi_R . 
\end{equation}
We write mass eigenvalues as $M_x \ (x=1,..,24)$. The orthogonal matrix $V_N$ and mass $M_x$ can be numerically obtained. 

{\it Charged fermion masses} :
we obtain Dirac mass terms of $e'_{1,2}$ from $M_{L'} {\rm Tr}[\bar L' L']$ term such that
\begin{equation}
M_{L'} {\rm Tr}[\bar L' L'] \supset M_{L'} (\overline{e'_1} e'_1 + \overline{e'_2} e'_2),
\end{equation} 
where generation index is omitted.
We choose the basis in which $M_{L'}$ is diagonal without loss of generality.

\section{Phenomenology}

In this section we discuss phenomenology of our model such as active neutrino mass generation, 
lepton flavor violation(LFV) and DM physics.

\subsection{Neutrino mass generation and LFV}

\begin{figure}[t]
\begin{center}
\includegraphics[width=80mm]{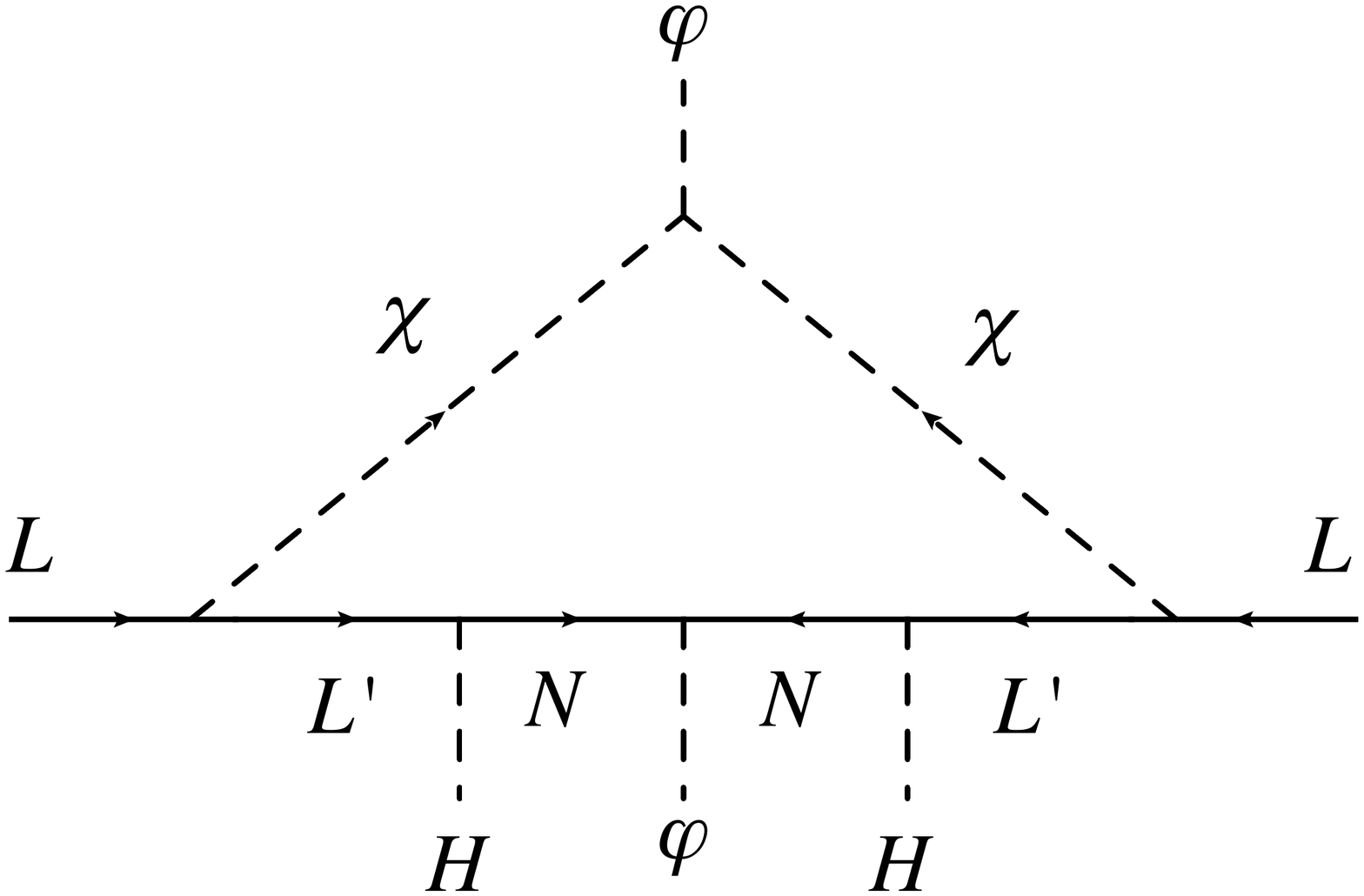} 
\includegraphics[width=80mm]{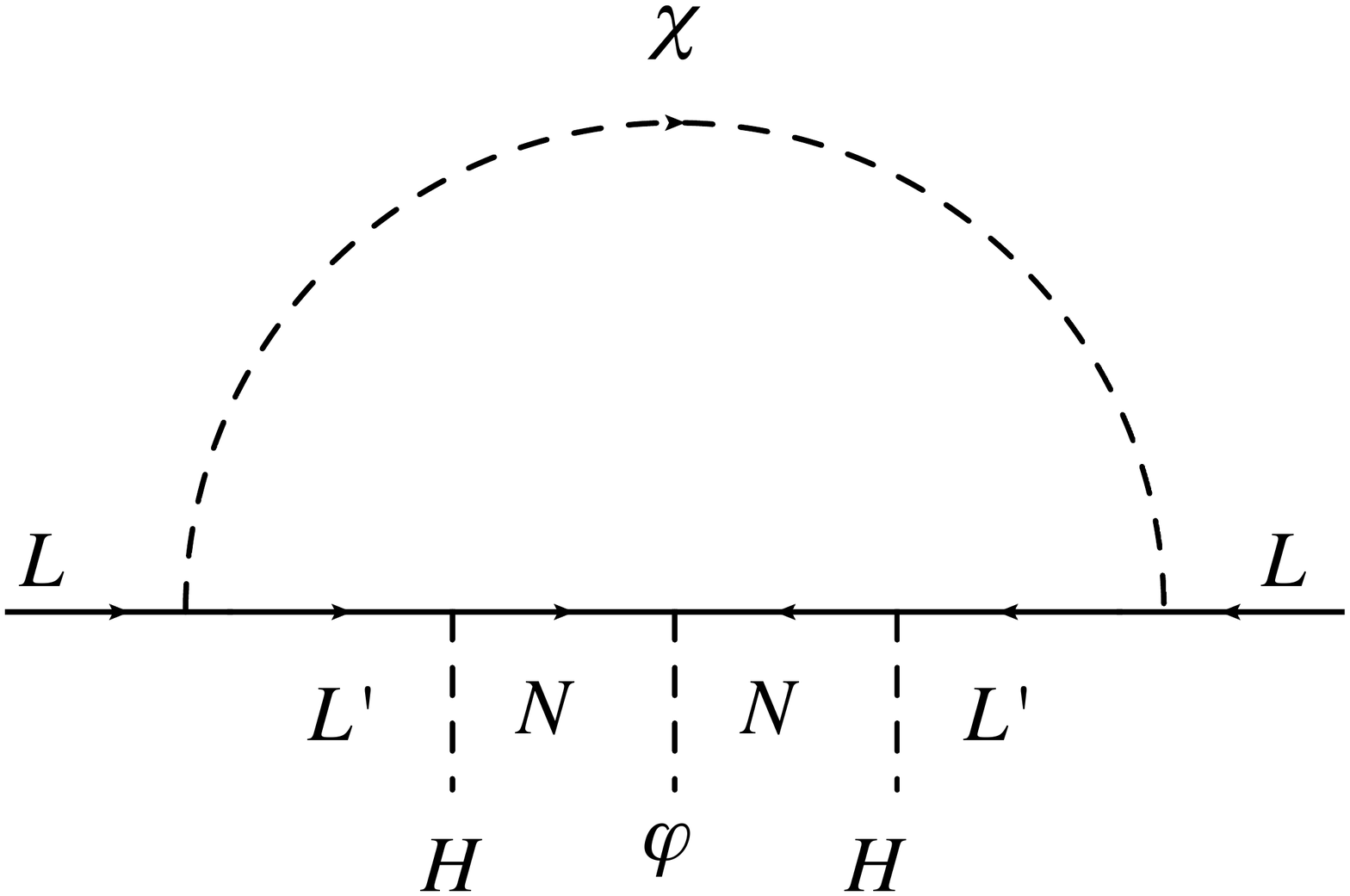} 
\caption{One-loop diagrams generating neutrino masses. } 
 \label{fig:neutrino}
\end{center}\end{figure}

In this subsection we discuss neutrino mass generation mechanism and related LFV processes.
The relevant interactions are obtained from the first two terms of the third line in RHS of Eq.~\eqref{eq:Lnew} 
Writing the terms by mass eigenstates, we obtain
\begin{align}
\mathcal{L} \supset & \frac{1}{\sqrt{2}} f^+_{ia} (V_{N})_{ax} \overline{\nu^i_L} \psi_R^x \rho_1^*  +  \frac{1}{\sqrt{2}} f^-_{ia} (V_{N})_{ax} \overline{\nu^i_L} \psi_R^x \rho_2^* \nonumber \\
& - \frac{1}{\sqrt{2}} f^-_{ia} (V_{N})_{3+a, x} \overline{\nu^i_L} \psi_R^x \rho_1 +  \frac{1}{\sqrt{2}} f^+_{ia} (V_{N})_{3+a, x} \overline{\nu^i_L} \psi_R^x \rho_2   \nonumber \\
& + \frac{1}{\sqrt{2}} f^+_{ia}  \overline{e^i_L} e'^a_{1R} \rho_1^* + \frac{1}{\sqrt{2}} f^-_{ia}  \overline{e^i_L} e'^a_{1R} \rho_2^*
- \frac{1}{\sqrt{2}} f^-_{ia}  \overline{e^i_L} e'^a_{2R} \rho_1 + \frac{1}{\sqrt{2}} f^+_{ia}  \overline{e^i_L} e'^a_{2R} \rho_2 + h.c.\, ,
\end{align}
where $f^+_{ia} = (f+f')_{ia}$ and $f^-_{ia} = (f-f')_{ia}$.

{\it Neutrino mass generation}: 
In our model active neutrino masses are generated through one-loop diagrams in Fig.~\ref{fig:neutrino} which is given by original flavor eigenstates in the Lagrangian.
The neutrino mass matrix is then calculated as
\begin{align}
(M_\nu)_{ij} &=   \frac{M_x }{(4\pi)^2} (f^+_{ia} f^-_{jb} m_{1}^2 - f^-_{ia}f^+_{jb} m_{2}^2) (V_N)_{ax} (V_N)_{3+b,x}  \int_0^1 [dX]_3 \frac{1}{x M_x^2 + y m_{1}^2+ z m_{2}^2} \nonumber \\
& \simeq  10^{-10} {\rm GeV} \frac{M_x}{\rm TeV}  \frac{\left(f^+_{ia} f^-_{jb} \frac{m_{1}^2}{m_{2}^2} - f^-_{ia}f^+_{jb}  \right) (V_N)_{ax} (V_N)_{3+b,x}}{10^{-11}}
\int_0^1 [dX]_3 \frac{m_{\rho_2}^2}{x M_x^2 + y m_{1}^2+ z m_{2}^2},
\end{align}
where $\int_0^1 [dX]_3 = \int_0^1 dx dy dz \delta(1-x-y-z)$.
It is possible to accommodate neutrino measurement by tuning the Yukawa couplings $f_{ia}$ where 
its magnitudes will be less than $\mathcal{O}(10^{-4})-\mathcal{O}(10^{-5})$ if value of $V_N$ components are around $0.1$ and $M_x={\cal O}(1)$ TeV.

{\it LFV decays of charged leptons}:
Yukawa interactions associated with charged dark fermions induce LFV decay of $\ell \to \ell' \gamma$ at one-loop level.
We then estimate the branching ratios by calculating relevant one-loop diagrams and
the branching ratio of $\ell_i \to\ell_j \gamma$ 
is given by
\begin{align}
B(\ell_i \to\ell_j \gamma)
=
\frac{48\pi^3 C_{ij} \alpha_{\rm em}}{{\rm G_F^2} m_i^2 }(|(a_R)_{ij}|^2+|(a_L)_{ij}|^2),
\end{align}
where $m_{i(j)},\ (i(j)=1,2,3)$ is the mass for the initial(final) eigenstate of charged-lepton identified as $1\equiv e,2\equiv\mu,3\equiv\tau$, and $(C_{21}, C_{31}, C_{32})=(1,0.1784, 0.1736)$.
Relevant amplitudes are estimated as
\begin{align}
(a_R)_{ij} = & - \sum_{K=1,2} \frac{f^+_{ja} f^{+ \dagger}_{ai}}{2(4\pi)^2} m_{\ell_j} \int_0^a [dX]_3 \frac{xz}{x(x-1) m_{\ell_i}^2 + x m_{K}^2 + (y+z) m_{e'^a_{K} }^2 } \nonumber \\
&  - \frac{f^-_{ja} f^{- \dagger}_{ai}}{2(4\pi)^2} m_{\ell_j} \int_0^a [dX]_3 \frac{xz}{x(x-1) m_{\ell_i}^2 + x m_{2}^2 + (y+z) m_{e'^a_{1} }^2 } \nonumber \\  
&   - \frac{f^-_{ja} f^{- \dagger}_{ai}}{2(4\pi)^2} m_{\ell_j} \int_0^a [dX]_3 \frac{xz}{x(x-1) m_{\ell_i}^2 + x m_{1}^2 + (y+z) m_{e'^a_{2} }^2 }  \\
(a_L)_{ij} = & - \sum_{K=1,2} \frac{f^+_{ja} f^{+ \dagger}_{ai}}{2(4\pi)^2} m_{\ell_i} \int_0^a [dX]_3 \frac{xy}{x(x-1) m_{\ell_i}^2 + x m_{K}^2 + (y+z) m_{e'^a_{K} }^2 } \nonumber \\
&  - \frac{f^-_{ja} f^{- \dagger}_{ai}}{2(4\pi)^2} m_{\ell_i} \int_0^a [dX]_3 \frac{xy}{x(x-1) m_{\ell_i}^2 + x m_{2}^2 + (y+z) m_{e'^a_{1} }^2 } \nonumber \\  
&   - \frac{f^-_{ja} f^{- \dagger}_{ai}}{2(4\pi)^2} m_{\ell_i} \int_0^a [dX]_3 \frac{xy}{x(x-1) m_{\ell_i}^2 + x m_{1}^2 + (y+z) m_{e'^a_{2} }^2 }.
\end{align}
The current experimental upper bounds for the BRs are given 
by~\cite{TheMEG:2016wtm, Aubert:2009ag,Renga:2018fpd}
  \begin{align}
  B(\mu\rightarrow e\gamma) &\leq4.2\times10^{-13},\quad 
  B(\tau\rightarrow \mu\gamma)\leq4.4\times10^{-8}, \quad  
  B(\tau\rightarrow e\gamma) \leq3.3\times10^{-8}~.
 \label{expLFV}
 \end{align}
 We can easily avoid these constraints when Yukawa coupling $f_{ia}$ and $f'_{ia}$ are smaller than $\mathcal{O}(10^{-3})$
 and it can be consistent with neutrino mass scale as discussed above.
Here we do not carry out explicit numerical analysis since we have sufficient degrees of freedom to realize neutrino oscillation data 
and LFV constraints can be easily avoided.

\subsection{Dark matter}

In our model DM candidates are $Z_4$ charged particles in dark sector which are $X^\pm$, $\rho_{1,2}$, $\varphi^\pm$ and $\psi^x_R$.
Here we consider a scenario in which $\rho_{1,2}$ and/or $X^\pm$ are DM candidates choosing the other particles to be heavier than them.
Under $Z_4$ symmetry, $\rho_{1,2}$ and $X^\pm$ transform as $\rho_{1,2} \to i \rho_{1,2}$ and $X^\pm \to - X^\pm$ respectively.
In our analysis we focus on gauge interactions of DM candidates since scalar portal interaction is preferred to be small to avoid constraints from direct detection experiments of DM
and Yukawa interaction $f_{ia}$ should be very small to realize neutrino mass as we discussed above.

Firstly interactions among dark gauge bosons are written by
\begin{equation}
\mathcal{L} \supset - g_X \epsilon^{\alpha \beta \gamma} \partial_\mu \tilde X^\alpha_\nu \tilde X^{\beta \mu} \tilde X^{\gamma \nu} 
- \frac14 g_X^2 \epsilon^{\alpha \beta \gamma} \epsilon^{\alpha \rho \sigma} \tilde X^\beta_\mu \tilde X^\gamma_\nu \tilde X^{\sigma \mu} \tilde X^{\rho \nu},
\label{eq:gauge_int}
\end{equation}
where $\epsilon^{\alpha \beta \gamma}$ is the structure constants of $SU(2)_D$.
We note that the four point gauge interaction in Eq.~(\ref{eq:gauge_int}) does not provide 
dominant contribution to DM annihilation process since $m_{Z'} \simeq m_{X^3} \gtrsim m_{X^\pm}$ in our scenario. 
Thus we focus on the three point interactions which are written in terms of mass eigenstates $X^\pm$ and $Z'$ such that 
\begin{align}
\mathcal{L} & \supset i g_X C_{ZZ'} \Bigl[ (\partial_\mu X^+_\nu - \partial_\nu X^+_\mu) X^{- \mu} Z'^\nu - (\partial_\mu X^-_\nu - \partial_\nu X^-_\mu) X^{+ \mu} Z'^\nu \nonumber \\
& \qquad \qquad \quad + \frac{1}{2} (\partial_\mu Z'_\nu - \partial_\nu Z'_\mu) (X^{+\mu } X^{- \nu} - X^{- \mu} X^{+ \nu}) \Bigr],
\end{align}
where $C_{ZZ'} \equiv \cos \theta_{ZZ'}$.
Through the $Z$--$Z'$ mixing, $Z'$ interactions with SM fermions $f$ are obtained as 
\begin{equation}
\mathcal{L}_{Z' \bar f f} = \sum_{X=L,R} \frac{g}{\cos \theta_W} Z'_\mu \bar f_X \gamma^\mu \left[  S_{ZZ'} (T_3 - Q_f \sin^2 \theta_W) + C_{ZZ'} \chi Y  \sin \theta_W  \right]   f_X,
\end{equation} 
where $T_3$ is diagonal generator of $SU(2)_L$, $Q_f$ is the electric charge of a SM fermion, and $S_{ZZ'} \equiv \sin \theta_{ZZ'}$.
Gauge interactions associated with scalar DM candidates $\rho_{1,2}$ are also obtained as 
\begin{align}
\mathcal{L} & \supset i \frac{g_X C_{ZZ'}}{2} Z'_\mu (\rho_1^* \partial^\mu \rho_1 - \rho_1 \partial^\mu \rho_1^* + \rho_2^* \partial^\mu \rho_2 - \rho_2 \partial^\mu \rho_2^*) \nonumber \\
& + i \frac{g_X}{\sqrt{2}} X^+_\mu (\rho^*_1 \partial^\mu \rho_2^* - \rho_2^* \partial^\mu \rho_1^*) - i \frac{g_X}{\sqrt{2}} X^-_\mu (\rho_1 \partial^\mu \rho_2 - \rho_2 \partial^\mu \rho_1) \nonumber \\
& + \frac{g_X^2}{4} C_{ZZ'}^2 Z'_\mu Z'^\mu (\rho_1^* \rho_1 + \rho_2^* \rho_2) - \frac{g_X^2}{2}  (X^+_\mu Z'^{\mu} \rho_1^* \rho^*_2 + X^-_\mu Z'^{\mu} \rho_1 \rho_2).
\end{align}
Note that our DM candidates can also interact with SM $Z$ boson through mixing effect.
These are obtained just substituting $\{C_{ZZ'}, Z'_\mu \}$ into $\{S_{ZZ'}, Z_\mu \}$. 
Such interactions are thus suppressed by tiny $S_{ZZ'}$ in our scenario.

\begin{figure}[t]
\begin{center}
\includegraphics[width=120mm]{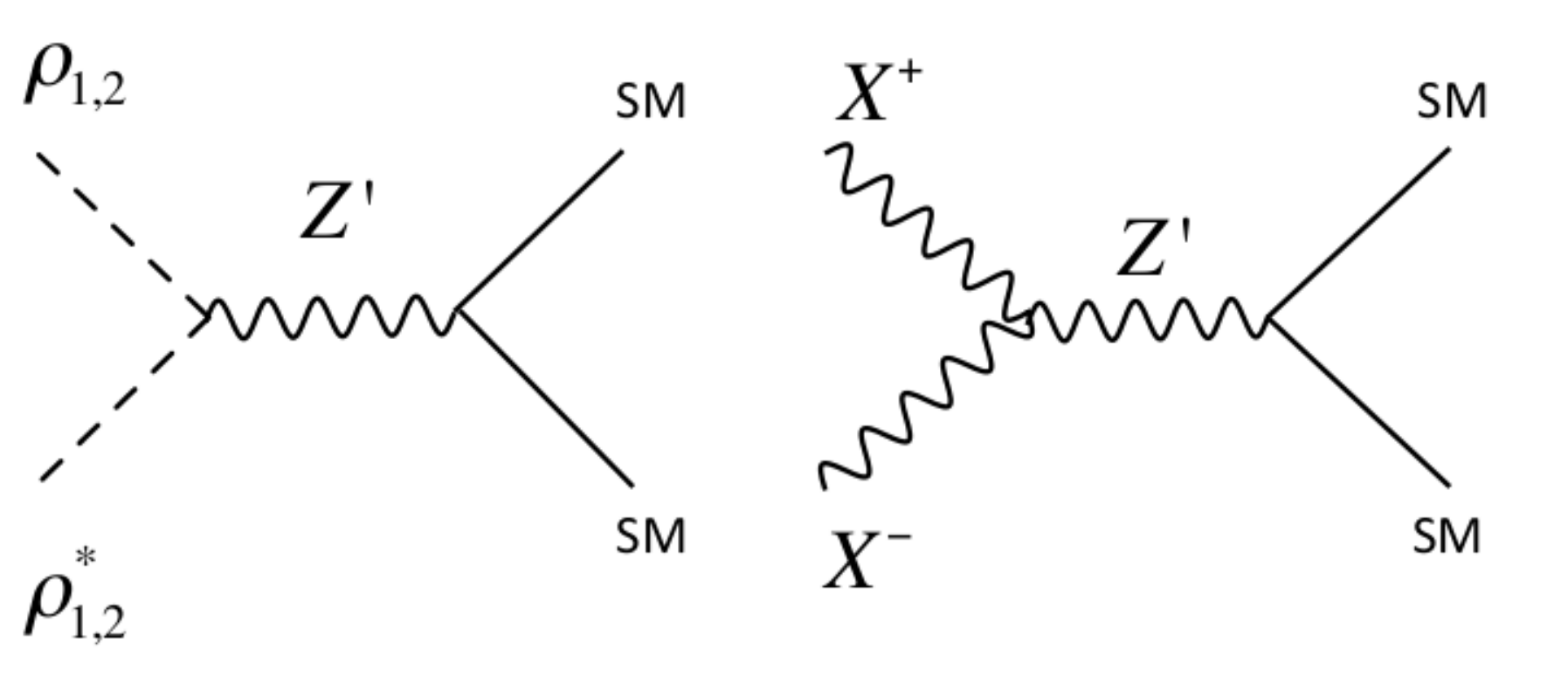}
\includegraphics[width=120mm]{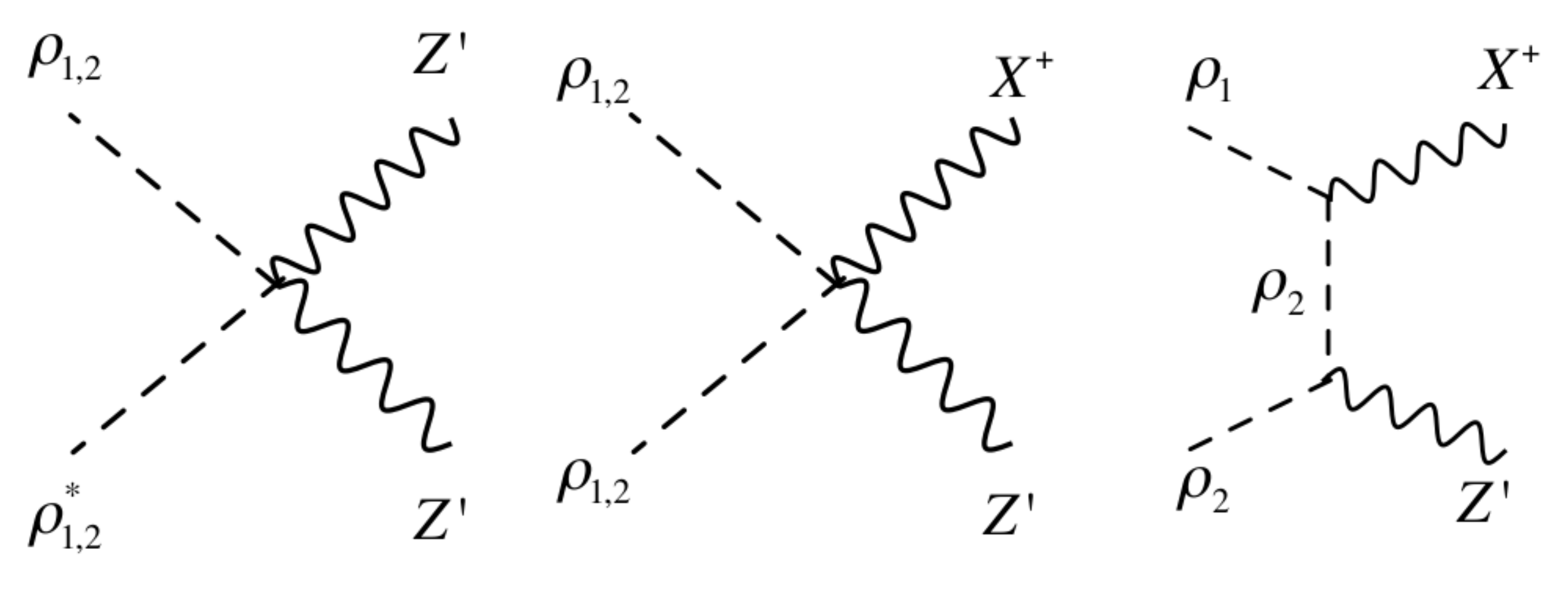} 
\caption{Feynman diagrams for DM annihilation processes. } 
 \label{fig:annihilation}
\end{center}\end{figure}

\begin{figure}[t]
\begin{center}
\includegraphics[width=80mm]{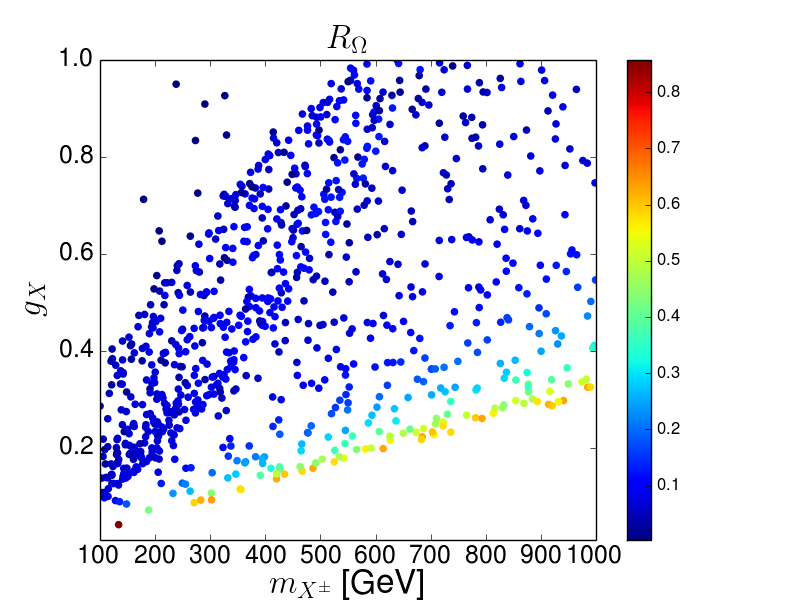} 
\includegraphics[width=80mm]{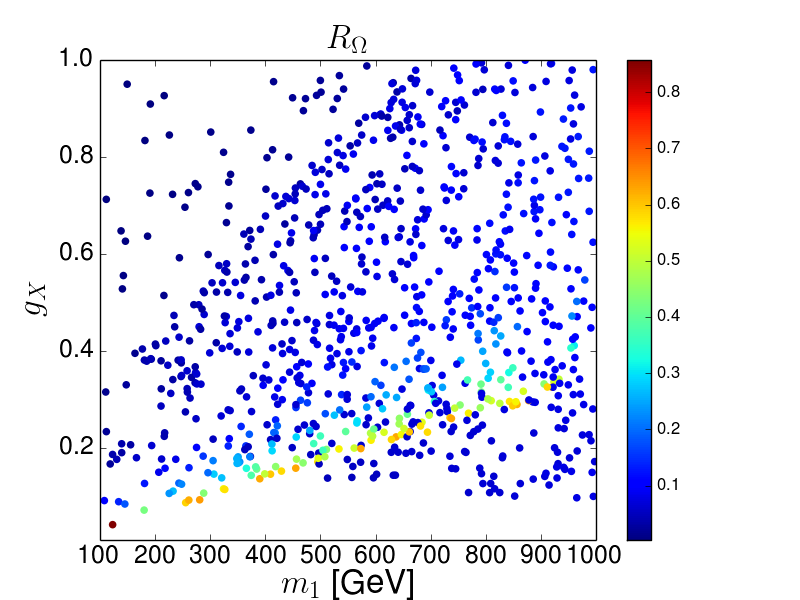} 
\caption{Parameter points satisfying observed relic density where color gradient indicates ratio of relic density $R_\Omega \equiv \Omega_{X^\pm}/\Omega_{\rho_1}$. 
For left and right panel, horizontal axis corresponds to mass of $X^\pm$ and $\rho_1$ respectively.} 
 \label{fig:DM1}
\end{center}\end{figure}

\begin{figure}[t]
\begin{center}
\includegraphics[width=80mm]{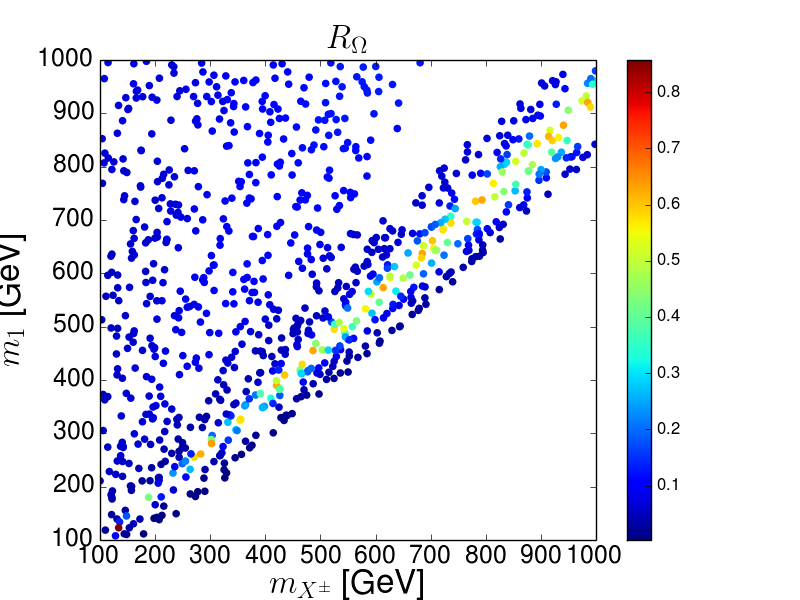} 
\caption{Parameter points satisfying observed relic density on two DM mass plane where color gradient is the same as Fig.~\ref{fig:DM1}.} 
 \label{fig:DM2}
\end{center}\end{figure}

In our scenario we have more than one DM components depending on mass relation among $X^\pm$, $\rho_1$ and $\rho_2$ as follows;
\begin{enumerate}
\item $m_{X^\pm} + m_{1} < m_{2} \ \rightarrow$  $X^\pm$ and $\rho_1$ are DM,  
\item $m_{1} + m_{2} < m_{X^\pm} \ \rightarrow$  $\rho_1$ and $\rho_2$ are DM,  
\item $m_{2} - m_{1} < m_{X^\pm} < m_{1} + m_{2} \ \rightarrow$ $\rho_1$, $\rho_2$ and $X^\pm$ are DM.
\end{enumerate}
Here $\rho_2$ decays into $\rho_1^* X_+$ in case 1 while $X^\pm$ decays into $\rho_1^{(*)} \rho_2^{(*)}$ in case 2.
In the following, we consider the case 1 above; in case 2 scalar portal interaction tends to be required as dark gauge bosons are heavier and case 3 is more complicated as we have three DM components.
In case 1 relevant DM annihilation processes are given by diagrams in Fig.~\ref{fig:annihilation};
pair annihilation of $\rho_{1,2}$, pair annihilation of $X^\pm$, semi-annihilation $\rho_{1(2)} \rho_{1(2)} \to X^+ Z'$, and semi-coannihilation $\rho_1 \rho_2 \to X^+ Z'$ ($Z'$ decays into SM particles).
Then we scan relevant parameters $\{m_{X^\pm}, m_{1}, g_X\}$ within the following region
\begin{equation}
m_{X^\pm} \in [100, 1000] \ {\rm GeV}, \quad m_{1}  \in [100, 1000] \ {\rm GeV}, \quad g_X \in [0.01, 1.0],
\end{equation}
where we fix the other parameters as $m_{2} = m_{X^\pm} + m_{1} + 100$ GeV, $v_\varphi^2/v_\Phi^2 = 0.1$ and $\chi = 5 \times 10^{-4}$.
We then estimate the relic density of $\rho_1$ and $X^\pm$ with {\it micrOMEGAs 5}~\cite{Belanger:2014vza} implementing relevant interactions.
In left and right panels of Fig.~\ref{fig:DM1}, we show parameter points on $\{m_{X^\pm}, g_x \}$ and $\{m_{1}, g_x \}$ planes satisfying observed relic density in approximated region as $0.11 < \Omega h^2 = (\Omega_{X^\pm}+ \Omega_{\rho_1}) h^2 < 0.13$ around $\Omega h^2 \simeq 0.12$~\cite{pdg} where color gradient indicates ratio of the relic density for two DM components: $R_\Omega \equiv \Omega_{X^\pm}/\Omega_{\rho_1}$.  
In addition, we show the parameter points realizing observed relic density on $\{m_{X^\pm}, m_{1} \}$ plane in Fig.~\ref{fig:DM2}.
We find that relic density of $X^\pm$ tends to be smaller than that of $\rho_1$ since cross section of $X^\pm X^\pm \to Z' \to f_{SM} f_{SM}$ process is enhanced by resonant condition $m_{Z'} \sim 2 m_{X^\pm}$. 
Relic density of two components can be similar order when $m_{X^\pm} \sim m_{1} \sim m_{Z'}/2$.
Note also that relic density can be explained in the region where $\rho_1$ is much heavier than $X^\pm$.
In this region $\rho_{1(2)}$ can annihilate into dark gauge bosons including semi-annihilation and semi-coannihilation processes and relic density of $\rho_1$ can be reduced to satisfy $\Omega h^2 \sim 0.12$.
Here we comment on constraints from direct detection of our DM candidates.
Our DM can interact with nucleon via $Z'$ and $Z$ boson exchange processes associated with $Z$--$Z'$ mixing.
In our model, the mixing is small since it is induced at loop level and DM-nucleon scattering cross section will be sufficiently small to avoid current constraints.

\section{Summary and discussion}

We have discussed a model based on dark $SU(2)_D$ gauge symmetry in which we introduce 
$SU(2)_L \times SU(2)_D$ bi-doublet vector like leptons.
The bi-doublets connect dark sector and SM sector through the interaction associated with the SM lepton doublets and $SU(2)_D$ scalar doublet.
We then obtain active neutrino masses and an interaction realizing kinetic mixing between $SU(2)_D$ and $U(1)_Y$ gauge fields at loop level.
Moreover there is remnant $Z_4$ symmetry after spontaneous breaking of $SU(2)_D$ in our scenario and the symmetry guarantees the stability of our DM candidates.

We have formulated active neutrino mass matrix and related LFV processes in our model.
Also relic density of our DM candidates is estimated scanning some relevant parameters.
Remarkably we have found multicomponent scenario in some parameter space where we have vector and scalar DM components.
We have shown parameter points realizing observed relic density where vector DM tends to provide smaller relic density due to resonant enhancement of corresponding annihilation cross section.

\section*{Acknowledgments}
\vspace{0.5cm}
{\it
This research was supported by an appointment to the JRG Program at the APCTP through the Science and Technology Promotion Fund and Lottery Fund of the Korean Government. This was also supported by the Korean Local Governments - Gyeongsangbuk-do Province and Pohang City (H.O.). H. O. is sincerely grateful for the KIAS member.}


\appendix
\section{Some formula for $SU(2)_D$ quintet}

Here we summarize some formula to write interactions for $SU(2)_D$ quintet.
We write $SU(2)_D$ generators in $5 \times 5$ form denoted by ${\cal T}^{(5)}_a$ such that 
\begin{eqnarray} & \displaystyle
{\cal T}_1^{(5)} \,\,=\,\, \frac{1}{2}
\begin{pmatrix}
0 & 2 & 0 & 0 & 0 \\
2 & 0 & \sqrt{6} & 0 & 0 \\
0 & \sqrt{6} & 0 & \sqrt{6} & 0 \\
0 & 0 & \sqrt{6} & 0 & 2 \\
0 & 0 & 0 & 2 & 0
\end{pmatrix} , \hspace{5ex}
{\cal T}_2^{(5)} \,\,=\,\, \frac{i}{2}
\begin{pmatrix}
0 & -2 & 0 & 0 & 0 \\
2 & 0 & -\sqrt{6} & 0 & 0 \\
0 & \sqrt{6} & 0 & -\sqrt{6} & 0 \\
0 & 0 & \sqrt{6} & 0 & -2 \\
0 & 0 & 0 & 2 & 0
\end{pmatrix} ,
& \nonumber \\ & \displaystyle
{\cal T}_3^{(5)} \,\,=\,\, {\rm diag}(2,1,0,-1,-2) ~.
\end{eqnarray}
We can then write $SU(2)_D$ gauge interactions for quintet $\Phi$ by kinetic term $(D_\mu \Phi)^\dagger (D^\mu \Phi)$ where covariant derivative is
\begin{align} 
D_\mu \Phi = (\partial_\mu + i g_X {\cal T}_\alpha^{(5)} X_\mu^\alpha ) \Phi_5.
\end{align}

\end{document}